\documentclass[twocolumn, pra, superscriptaddress,notitlepage, nofootinbib]{revtex4-2}
\usepackage{amsmath}
\usepackage{graphicx}
                                                
\usepackage{mathtools}
\usepackage{ytableau}
\usepackage{amssymb,amsmath,amstext,amsthm}
\usepackage{graphicx}
\usepackage{epstopdf}
\usepackage{color}
\usepackage{dsfont}
\usepackage{bm}
\usepackage{appendix}
\usepackage[T1]{fontenc}
\usepackage{bbm}
\usepackage{latexsym}
\usepackage{bbm}
\usepackage{verbatim}
\usepackage{lipsum}
\usepackage{braket}
\usepackage{ulem}
\usepackage{mathtools}
\usepackage{mathrsfs}
\usepackage{dcolumn}
\usepackage{xcolor}   
\usepackage{bm}        
\usepackage{tikz}
\usepackage{gensymb}
\usepackage{xcolor}
\usepackage{float}
\usepackage{dsfont}
\usepackage[caption = false]{subfig}

 \usepackage[english]{babel}
 \usepackage[bookmarks,bookmarksopen,bookmarksdepth=2]{hyperref}
\hypersetup{
    colorlinks=true,
    citecolor=blue,
    linkcolor=blue,
    filecolor=magenta,
    urlcolor=blue,
}
 \usepackage{quantikz}
\usepackage[capitalize]{cleveref}
\usepackage{xcolor}
\usepackage{colortbl}

\DeclarePairedDelimiter\floor{\lfloor}{\rfloor}

\usepackage{color}
\usepackage{booktabs}
\usepackage{caption}
\usepackage{subcaption}
\PassOptionsToPackage{table}{xcolor}
\usepackage{algorithm}
\usepackage{algorithmic}

\renewcommand{\arraystretch}{1.56}
\begin{document}
\def \addJPMorgan{Global Technology Applied Research, JPMorgan Chase, New York, NY 10017, USA}

\title{New Improvements in Solving Large LABS Instances Using Massively Parallelizable \\Memetic Tabu Search}

\author{Zhiwei Zhang}
\affiliation{\addJPMorgan}
\author{Jiayu Shen}
\affiliation{\addJPMorgan}
\author{Niraj Kumar}
\thanks{Corresponding Author.\\Email: \url{niraj.x7.kumar@jpmchase.com}}
\affiliation{\addJPMorgan}
\author{Marco Pistoia}
\thanks{Principal Investigator.\\Email: \url{marco.pistoia@jpmchase.com}}
\affiliation{\addJPMorgan}

\date{\today}

\begin{abstract}
Low Autocorrelation Binary Sequences (LABS) is a particularly  challenging binary optimization problem which quickly becomes intractable in finding the global optimum for problem sizes beyond 66. This aspect makes LABS appealing to use as a test-bed for meta-heuristic optimization solvers to target large problem sizes. In this work, we introduce a massively parallelized implementation of the memetic tabu search algorithm to tackle LABS problem for sizes up to 120. By effectively combining the block level and thread level parallelism framework within a single Nvidia-A100 GPU, and creating hyper optimized binary-valued data structures for shared memory among the blocks, we showcase up to 26 fold speedup compared to the analogous 16-core CPU implementation. Our implementation has also enabled us to find new LABS merit factor values for sixteen different problem sizes between 92 and 120. Crucially, we also showcase improved values for five odd-sized problems $\{99, 107, 109, 113, 119\}$ whose previous best known results coincided with the provably optimal skew-symmetric search sequences. Consequently, our result highlights the importance of a focus on general-purpose solver to tackle LABS, since leveraging its skew-symmetry could lead to sub-optimal solutions.
\end{abstract}

\maketitle

\section*{Introduction}

Meta-heuristic methods, also known as local-neighborhood search methods, have become pivotal in the advancement of solving combinatorial optimization problems \cite{blum2003metaheuristics}. Methods such as evolutionary algorithms, simulated annealing, and tabu search, offer a flexible framework adaptable to a wide range of problems, often providing near-optimal solutions with reasonable computational time \cite{bartz2014evolutionary,back1993overview,van1987simulated,glover1990tabu}.  Among these meta-heuristics, the memetic tabu search has shown remarkable efficacy in navigating large and complex landscapes \cite{silva2021quadratic}. This hybrid approach combines the global search capabilities inherent in the memetic evolutionary algorithms with local stochastic search of tabu search method equipped with adaptive memory, to enhance solution quality and convergence speed. This technique has been applied to a variety of problems with great success; graph coloring~\cite{10.1007/978-3-540-75755-9_130}, vehicle routing problems~\cite{CORDEAU20122033}, traveling salesman problems~\cite{349949,FIECHTER1994243}, quadratic assignment problems~\cite{chakrapani1993massively,349949,JAMES2009810}, large optimization problems on hundreds of heterogeneous machines~\cite{Talbi1999,349949}, and many other problems. A  literature survey on memetic tabu can be found in \cite{wilson2022review}.

Meta-heuristics stand out from other optimization techniques due to a plethora of reasons. First, they are gradient-free methods and allow for moves based on acceptance-rejection criteria, which particularly helps avoid early-stopping to suboptimal solutions, as opposed to gradient descent based methods \cite{gallardo2007memetic}. Secondly they balance the exploration and exploitation of the candidate solutions in the search landscape, thus allowing for a more thorough examination of the search landscape. Finally, the parallelization of these techniques with random initial starts further amplifies their capability, enabling rapid search space exploration. By distributing the search mechanism across multiple processors or cores, parallel meta-heuristics, in particular memetic-tabu, efficiently explore diverse regions of the solution space, thereby increasing the likelihood of identifying high-quality solutions \cite{crainic2005parallel}.

Similar to meta-heuristics, global optimization methods such as branch-and-bound can also be employed to tackle these optimization problems with rigorous guarantees of finding the optimal solutions \cite{lawler1966branch, guide2020cuda}. However, for a variety of such large-scale problems, they can be computationally prohibitive due to their exhaustive search nature. In contrast, due to the versatility and parallelization of meta-heuristics, they often rapidly converge to high-quality solutions without the need for exhaustive enumeration, outperforming global solvers in terms of speed and scaling with the problem size.

In this work, we realize the massively parallelized memetic tabu search algorithm to solve the Low Autocorrelation Binary Sequence (LABS) problem \cite{Boehmer1967}. Characterized by long-range fourth-order spin interaction terms within its objective function, LABS presents substantial computational challenges in finding solutions with minimal objective function values, even for relatively modest problem sizes, due to its non-linear and non-convex nature. Consequently, it serves as an excellent testbed for evaluating the performance of optimization solvers, particularly those based on meta-heuristic approaches. The LABS problem is significant in various fields, including telecommunications and physics, where it is used to design sequences with minimal autocorrelation \cite{Boehmer1967,Schroeder1970,Golay1977,Bernasconi1987}. 

\begin{figure*}[t!]
    \centering
    \includegraphics[width=1\linewidth]{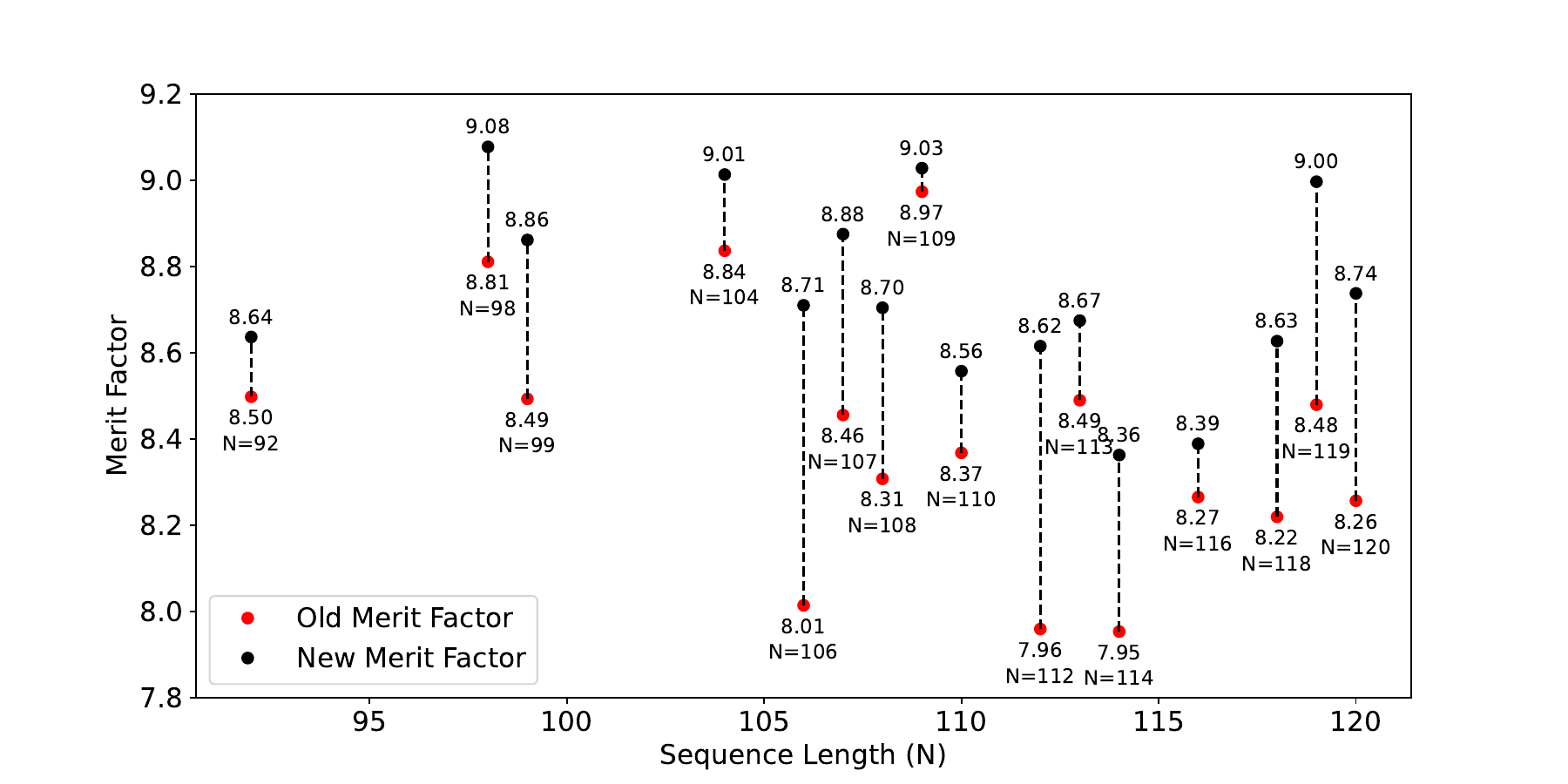}
    \caption{Improvements in the LABS merit factor objective value using our massively GPU parallelized memetic tabu search for sixteen new problem sizes between $N=92$ to $N=120$, improving the MF range from $[7.95, 8.97]$ to $[8.36,9.08]$. The new sequences for odd $N=99,107,109,113,119$ improve the MF from optimal skew-symmetric sequences.
    }
    \label{fig:improvement}
\end{figure*}

Memetic tabu search is known to particularly exhibit a competitive runtime scaling for LABS~\cite{gallardo2007memetic}.  Our efficient GPU parallelization builds upon this technique to allow for finding best known LABS solutions upto problem size 120 using a single Nvidia-A100 GPU with 6912 CUDA cores, 80 GB memory and 164 KB shared memory per streaming processor, within 96 hours for odd problem sizes $\leqslant 109$ and even problem sizes $\leqslant 118$. Crucially, our approach has enabled finding new LABS sequences for sixteen different problem sizes between 92 and 120 with better objective value compared to any reported previous solutions \cite{OPUS2-git_labs-Boskovic}. Moreover, for five of these odd problem sizes, our improved objective values surpass those achieved by the provably optimal skew-symmetric sequences, which were previously the best reported in the literature \cite{Bokovi2017}. Our findings suggest that relying solely on the skew-symmetric property of LABS to constrain the search space may result in sub-optimal solutions. Consequently, this work highlights the importance of enhancing the general-purpose solvers to effectively address this particularly challenging problem. 

The following sections are structured as follows. Section~\ref{sec:summary} provides a succinct summary of our key results. We describe the LABS problem along with its hardness, underlying symmetry, and existing methods in Section~\ref{sec:problem-description}. The memetic tabu methodology followed by our GPU parallelization technique is discussed in Section~\ref{sec:methodology} and \ref{sec:gpu-parallelism}. We detail our results, including the new MF values, scaling, and deviation from skew-symmetry in Section~\ref{sec:results}. We conclude of our work in Section~\ref{sec:discussion}.

\section{Summary of Results} \label{sec:summary}

The primary contributions of our work are as follows,

\begin{enumerate}
    \item A delicate design of a two-level ($block$ and $thread$) parallelism framework to allow for an optimized implementation of memetic tabu search for LABS problem, based on the embarrassing parallelization framework on GPUs. 
    \item Exploitation of fast shared memory in the Nvidia-A100 GPU by optimizing the algorithm's data structure storage into binary-value, enabling maximum GPU utility. This results in shared memory size of 5 KB, sufficient to run LABS for problem size up to 187. 
    \item Sixteen new LABS merit factor values found for both odd and even sequences with problem sizes between 92 and 120 (see Figure~\ref{fig:improvement} and Table~\ref{tab:labs_results}).
    \item Analysis and  evidence that our new sequences deviate from skew symmetric sequences, thus not likely to be found by first applying skew-symmetry and then unrestricted search. 
    \item GPU implementation achieving up to 26-fold speedup in time-to-solution compared to 16-core CPU implementation (Figure~\ref{fig:acceleration}). 
\end{enumerate}

\section{Problem Description} \label{sec:problem-description}

We begin by formally describing the LABS problem. Consider a binary sequence of length $N$, $S = s_1s_2 \cdots s_N$ with spin variable $s_i \in \{-1, 1\}$ for $1 \leqslant i \leqslant N$. Define the aperiodic autocorrelation of elements in sequence $S$ with distance $k$ as,
\begin{equation}
C_k(S) = \sum_{i=1}^{N-k} s_i s_{i+k}.
\end{equation}
The objective of LABS is to find the optimal assignment of $S$ which minimizes the corresponding energy function, expressed as the quadratic sum of its correlations,
\begin{equation}
S^* = \underset{S \in \{-1,1\}^N}{\text{arg min}} \hspace{1mm}E_N(S) = \sum_{k=1}^{N-1} C_k^2(S),
\label{eq:energy}
\end{equation}
or, equivalently, maximizes the merit factor (MF),
\begin{equation}
S^* = \underset{S \in \{-1,1\}^N}{\text{arg max}} \hspace{1mm}\mathcal{F}_N(S) = \frac{N^2}{2E_N(S)}
.
\end{equation}

\subsection{Hardness of LABS and Existing Methods}
\begin{figure*}[!htbp]
    \centering
    \includegraphics[width=1\linewidth]{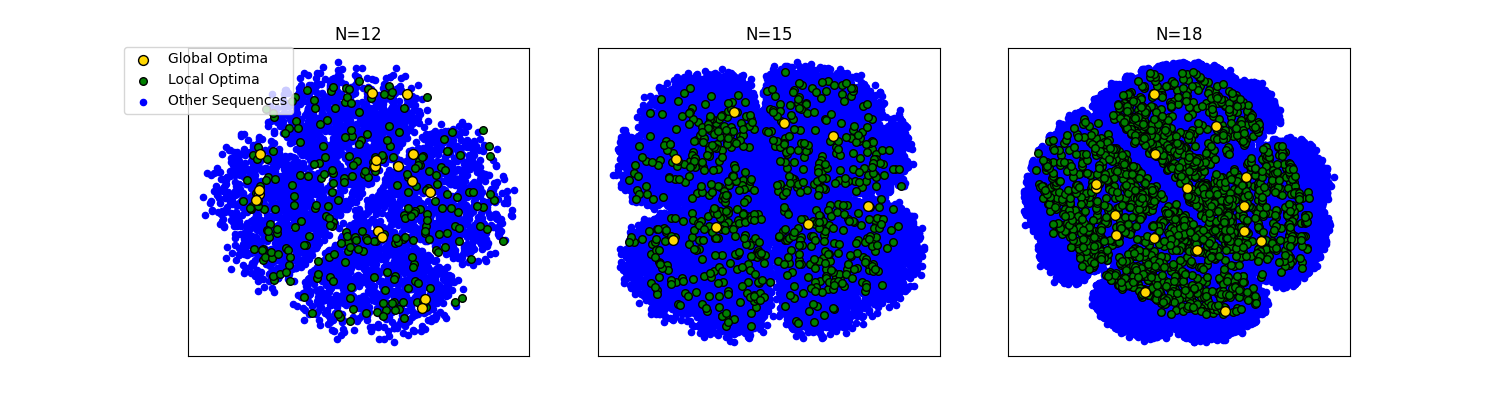}
    \caption{
   2-D projections of the solution spaces of LABS problem for $N=12,15,18$. All $2^N$ sequences are mapped by Uniform Manifold Approximation and Projection (UMAP) \cite{mcinnes2018umap} algorithms, approximately depicting the locality and global structure of the solution space. A sequence is marked a local optimum if its merit factor is strictly higher than all its $N$ neighbors. The number of local optima grows exponentially \cite{de1999metastable}, surrounding a limited number of isolated global optima, making LABS quickly intractable as $N$ grows.
    }
    \label{fig:landscape}
\end{figure*}
As a binary combinatorial optimization problem, LABS has a search space of $2^N$ for a problem of size $N$. The particular hardness of LABS comes due to the fourth order dependency among the spin variables $\{s_i, s_j, s_k, s_l\} \in S$ in the energy function, or, equivalently, the MF expression. As highlighted by the authors in \cite{de1999metastable}, the LABS energy function objective (and equivalently the MF) has $\mathcal{O}(N^2)$ non-zero quadratic and $\mathcal{O}(N^3)$ non-zero fourth order spin interactions. This leads to the search landscape of LABS being equivalent to a dilute 4-spin Ising glass model with exponentially (in $N$) many local minimas, while the global minima becomes extremely isolated deep and narrow resembling the shape of ``golf-holes''. We visually depict the spread of local and global minima for $N = 12,15,18$ in Figure~\ref{fig:landscape} to further illustrate this point. This explains the inability of any known method to reach the asymptotic merit factor of $\mathcal{F}_N \approx 12.3248$ as $N \rightarrow \infty$ as proposed by Golay using the ergodicity postulate \cite{Golay1982}\footnote{The largest known LABS experiments reported to date is for $N = 1010$ with $\mathcal{F}_N = 6.367$ which is a significant departure from Golay's postulate for large $N$.}.

With regard to provably optimal sequences, the branch-and-bound technique has been employed to find and prove the optimality of sequences for LABS with $N \leqslant 66$ \cite{Packebusch2016}. While it is a promising result, the reported runtime scaling of the above global optimization method is $\mathcal{O}(1.73^N)$ which quickly renders it infeasible for large $N$. For context, the wallclock time, as reported by \cite{Packebusch2016}, to solve $N=66$ was roughly 55 days using the Linux cluster comprising of a total of 248 (virtual) cores.

Simultaneously, a suite of mata-heuristic methods have also been applied to solve LABS in the past decades. While these methods do not provide theoretical guarantees of the optimality of the reached solution, they usually yield faster runtime given the target objective function value. 
Among such known general-purpose meta-heuristics for LABS, Kernighan–Lin~\cite{Brglez2003} exhibits a runtime scaling of $\mathcal{O}(1.463^N)$, the evolutionary-strategies algorithm~\cite{Brglez2003} exhibits $\mathcal{O}(1.397^N)$, and memetic tabu~\cite{gallardo2007memetic} exhibits $\mathcal{O}(1.34^N)$. In addition, the quantum approximate optimization algorithm combined with quantum minimum finding has been applied to solve LABS till problem size 40 with a scaling $\mathcal{O}(1.21^N)$ to demonstrate the evidence of scaling advantage for the quantum algorithm~\cite{Shaydulin2024}.

\subsection{Sequence Symmetries of LABS}

A simple observation would reveal that a LABS sequence has two exact symmetries \textit{i.e.}, these operations on the sequences preserve the objective function value exactly~\cite{Bokovi2017},
\begin{enumerate}
    \item \textit{complementation}: $s_i \mapsto -s_i$, $\forall i\in [N]$
    \item \textit{reversal}: $s_i \mapsto s_{N+1-i}$, $\forall i\in [N]$
\end{enumerate}
Respecting these symmetries while performing the optimization, results in a four fold reduction of the search space. 
We note that a straightforward method to implement the complementation symmetry is by setting the first spin variable in the sequence, $s_1\rightarrow 1$ which reduces the search space trivially by half. 

\subsection{Skew-Symmetry of LABS}
\label{subsec:skew_symmery}

Another class of restriction into the LABS sequence structure is the presence of skew-symmetric sequences for problems of odd length, as introduced by Golay \cite{Golay1972}. Formally, for an odd length $N=2n-1$, skew-symmetry is defined as,
\begin{equation}
    s_{n+i} = (-1)^i s_{n-i}, \hspace{2mm} i \in [n-1]
    .
\end{equation}
As shown by Ref.~\cite{dimitrov2022new}, sequences with this restriction result in the aperiodic autocorrelation values $C_k = 0$ for all odd $k$. 
The restriction over searching for solution within the skew-symmetric constraint significantly reduces the computational complexity of solving LABS due to fact that skew-symmetric subspace is of size $2^{(N+1)/2}$ as opposed to $2^N$ for unrestricted search. This has been exploited to find the optimal skew-symmetric solutions for odd $N \leqslant 89$ values using the branch-and-bound technique with a complexity $\mathcal{O}(1.34^N)$ \cite{Bokovi2017} \footnote{We also note that Ref.~\cite{Packebusch2016} reports optimal skew-symmetric sequences for odd $N \leqslant 119$, although they do not comment on the scaling factor.}, as opposed to $\mathcal{O}(1.73^N)$ exhibited by the method using unrestricted search \cite{Packebusch2016}. Furthermore, the integration of this constraint into meta-heuristic search methods has significantly enhanced their efficiency for odd $N$. Specifically, the memetic tabu search achieves a complexity scaling of $\mathcal{O}(1.16^N)$, as opposed to $\mathcal{O}(1.34^N)$ for an unrestricted search \cite{gallardo2007memetic, Bokovi2017}. Additionally, self-avoiding walks \cite{bovskovic2017low} demonstrate a scaling of $\mathcal{O}(1.15^N)$ \cite{Bokovi2017}, and the stochastic search method xLastovka exhibits a scaling of $\mathcal{O}(1.18^N)$ \cite{Brest2018}. Inspired by this idea, efforts have also been directed towards finding LABS sequences with improved objective values for even $N \gtrsim 150$, thorough concepts of  quasi-skew-symmetry and sequence operators \cite{dimitrov2022new}, and dual-step optimization that combines skew-symmetric search with unrestricted search \cite{pvsenivcnik2024dual}.

However, skew-symmetry is not necessarily an exact symmetry for optimal solutions of odd $N$, \textit{i.e.}, for some values of odd $N$, the optimal solutions do not exhibit this property. This is evident from the fact that out of the 32 odd values of $N$ within $3 \leqslant N \leqslant 66$, as reported by Ref.~\cite{Packebusch2016}, only 16 values of $N$ have skew-symmetric optimal solutions. Additionally, 6 values of $N$ have both skew-symmetric and non-skew-symmetric optimal solutions. Furthermore, crucially, 10 values of $N$ do not exhibit skew-symmetry in their optimal solutions. Hence, restricting the search space to incorporate this constraint, does not guarantee the optimal solution for odd $N$ values. 

In this work, we showcase that without imposing the skew-symmetry constraint, we find sequences for five odd $N = 99, 107, 109, 113, 119$ with an objective value better than the value obtained by optimal skew-symmetric sequences. Further, we analyze the results from Ref.~\cite{pvsenivcnik2024dual} to show that their newly found even-$N$ sequences using skew-symmetry search followed by unrestricted search, are still relatively close to a skew-symmetric sequence. This is in contrast to the large deviation from skew-symmetry in our new found sequences for even $N$'s with the general-purpose memetic tabu solver. This may imply that using neighborhoods of skew-symmetric sequences might make the search easier to be trapped, and a full-space search may still be useful, especially given the search landscape of the LABS problem.

\section{Methodology} \label{sec:methodology}
\label{sec:methodology}
To make this paper self-contained, we revisit the methodology of the memetic algorithm integrated with tabu search for addressing the LABS problem, as proposed in \cite{gallardo2007memetic}. In this section, we extend this approach by efficiently implementing the parallelization on GPUs. This hybrid approach combines the global search capabilities inherent in the memetic evolutionary algorithms \cite{bartz2014evolutionary} with local stochastic search of tabu search. The memetic algorithm operates by maintaining a population of binary sequences that are initialized randomly. During each iteration, it generates a new sequence, referred to as a child sequence, through probabilistic recombination and mutation processes. This child sequence subsequently serves as the initial point for the tabu search algorithm. Tabu search equips the greedy local search method with a tabu list to avoid flipping recently altered bits, thus facilitating a more thorough exploration of the search space \cite{glover1990tabu}. The combination of memetic and tabu search allows for effective exploration and exploitation of the solution space, balancing between diversification and intensification.

\subsection{Memetic Algorithms}

\begin{figure}[ht]\begin{algorithm}[H]
\begin{algorithmic}[1]
\STATE \texttt{population} $\gets$ $K$ random binary sequences of length $N$
\STATE \texttt{bestSeq} $\gets$ sequence with lowest energy from \texttt{population}
\WHILE{\texttt{E}(\texttt{bestSeq}) > \texttt{targetE}}
    \IF{$\texttt{randReal}(0,1) < p_{\text{comb}}$}
        \STATE $\texttt{parent}_1, \texttt{parent}_2 \gets \texttt{selectParents}(\texttt{population})$
        \STATE $\texttt{child} \gets \texttt{Combine}(\texttt{parent}_1, \texttt{parent}_2)$
    \ELSE
        \STATE $\texttt{child} \gets \texttt{Sample}(\texttt{population})$
    \ENDIF
    \STATE $\texttt{Mutate}(\texttt{child}, p_{\text{mutate}})$
    \STATE $\texttt{child} \gets \texttt{Tabu-Search}(\texttt{child})$
    \STATE update \texttt{bestSeq} to \texttt{child} if \texttt{E}(\texttt{bestSeq}) $>$ \texttt{E}(\texttt{child})
    \STATE replace a random individual in \texttt{population} by \texttt{child}
\ENDWHILE
\end{algorithmic}
\caption{\texttt{Memetic-Tabu}(N, \texttt{targetE})}\label{algo:memetictabu}
\end{algorithm}\end{figure}

The pseudo code for the memetic tabu algorithm used in this work is shown in Algorithm \ref{algo:memetictabu}. The algorithm starts by initializing the population of $K$ random binary sequences and creating a \texttt{bestSeq} register storing the sequence with the lowest energy among the chosen $K$ sequences. Subsequently in each iteration, with probability $p_{comb}$, the algorithm selects two parent sequences and recombines them to produce a child sequence. Next, each bit of the child sequence is mutated with probability $p_{mutate}$.  The resulting sequence is then sent to the tabu local search algorithm for further exploration which outputs the new child sequence. If the energy corresponding to this new child sequence is lower than the \texttt{bestSeq} energy, the \texttt{bestSeq} is updated. Finally, a random individual in the initial population is replaced by the new child sequence returned by the local search solver. For this work, we set memetic algorithm parameters to be $K=100$, $p_{comb}=0.9$ and $p_{mutate} = \frac{1}{N}$
which is consistent with Ref.~\cite{gallardo2007memetic}.

\subsection{Tabu Search}

\begin{figure}[h!]\begin{algorithm}[H]
\caption{\texttt{Tabu-Search}(\texttt{seq})}\label{alg:ts}
\begin{algorithmic}[1]
\STATE initialize \texttt{tabuList}$[1,\cdots,N]=0$ and \texttt{bestSeqTS}=\texttt{seq}
\STATE \texttt{pivot} $\gets$ \texttt{seq}
\FOR{$t =1,2,\cdots, \texttt{maxIter}$}
    \STATE  $
i \gets \mathop{\arg\min}\limits_{i=1,\ldots,N, \texttt{tabuList}[i]<t} \texttt{E}(\texttt{flip}(\texttt{pivot}, i))
$
    \STATE /* Evaluate energy of all $N$ neighborhoods of \texttt{pivot}. randomly choose one if there are multiple */
    \STATE $\texttt{pivot}\gets\texttt{flip}(\texttt{pivot,i})$ 
    \STATE \texttt{tabuList}[i] = $t+ \texttt{randomInt(minTabu, maxTabu)}$
    \STATE update \texttt{tableC} and \texttt{vectorC}
    \STATE update \texttt{bestSeqTS} to \texttt{pivot} if \texttt{E}(\texttt{bestSeqTS}) $>$ \texttt{E}(\texttt{\texttt{pivot}})
\ENDFOR
\RETURN \texttt{bestSeqTS}
\end{algorithmic}
\end{algorithm}\end{figure}

Tabu search takes a sequence from the memetic algorithm and runs a greedy local search on it with the restrictions given by the tabu list. The tabu list maps each coordinate to the future time represented as a number of iterations after which the coordinate would be allowed to get flipped. In each iteration, the algorithm checks energy values of all $N$ neighborhoods of the current search pivot and greedily selects the best bit to flip unless it is forbidden by the tabu list in this iteration\footnote{The $j^{th}$ neighborhood of a given sequence $S$ is the new sequence obtained by flipping the  $j^{th}$ variable while the other variables remain the same.}. After a flip is made, the bit is forbidden from getting changed again for a random number of iterations. The algorithm returns the sequence visited in all iterations with the lowest energy.  The details of the tabu search procedure is shown in Algorithm \ref{alg:ts}. The setting of hyperparameters in Ref.~\cite{gallardo2007memetic} is $\texttt{maxIter}=\texttt{randInt}[\frac{N}{2},\frac{3}{2}N]$, $\texttt{minTabu}=\floor{0.1\times \texttt{maxIter}}, \texttt{maxTabu}=\floor{0.12\times \texttt{maxIter}}$. 

\subsection{Data Structures for Efficient Energy Evaluation}
Naively computing the energy function value of a sequence according to Eq.~\ref{eq:energy} takes $\mathcal{O}(N^2)$ time and checking all neighborhoods at a pivot would take $\mathcal{O}(N^3)$, which can be computationally prohibitive. Several data structures have been proposed to make each neighborhood energy evaluation in $\mathcal{O}(N)$ and checking all neighborhood $\mathcal{O}(N^2)$. We use the data structures proposed in Ref.~\cite{gallardo2007memetic}, named \texttt{tableC} and \texttt{vectorC}, as shown in 
Table \ref{combinedTable}, in our framework. By maintaining those  data structures of the pivot sequence in tabu search, the energy of each neighborhood can be computed in linear time. Constructing the structures takes $\mathcal{O}(N^2)$ time while updating them after flipping a bit takes linear time as only one column and row needs to be changed. 

\begin{table}[h]
\centering
\renewcommand{\arraystretch}{1.5} % Adjust row height
\begin{tabular}{|ccccc|c} % Ensure the last column has a vertical line
    \multicolumn{4}{c}{\texttt{tableC}(S)} &  \multicolumn{2}{c}{\texttt{vectorC}(S)} \\ 
    \cline{1-4} \cline{6-6}% Add horizontal line only under tableC columns
    $s_1s_2$ & $s_2s_3$ & $s_3s_4$ & $s_4s_5$ &  & $s_1s_2 + s_2s_3 + s_3s_4 + s_4s_5$ \\ 
    $s_1s_3$ & $s_2s_4$ & $s_3s_5$ & &  & $s_1s_3 + s_2s_4 + s_3s_5$ \\ 
    $s_1s_4$ & $s_2s_5$ & & & & $s_1s_4 + s_2s_5$ \\ 
    $s_1s_5$ & & & & & $s_1s_5$ \\
\end{tabular}
\vspace{0.5cm} % Add vertical space between tables
\caption{The binary, left upper triangle matrix \texttt{tableC} and the integer vector \texttt{vectorC}.}
\label{combinedTable}
\end{table}

\section{Massive Parallelism with GPU} \label{sec:gpu-parallelism}
\begin{figure*}[ht!]
    \centering
    \includegraphics[width=1\linewidth]{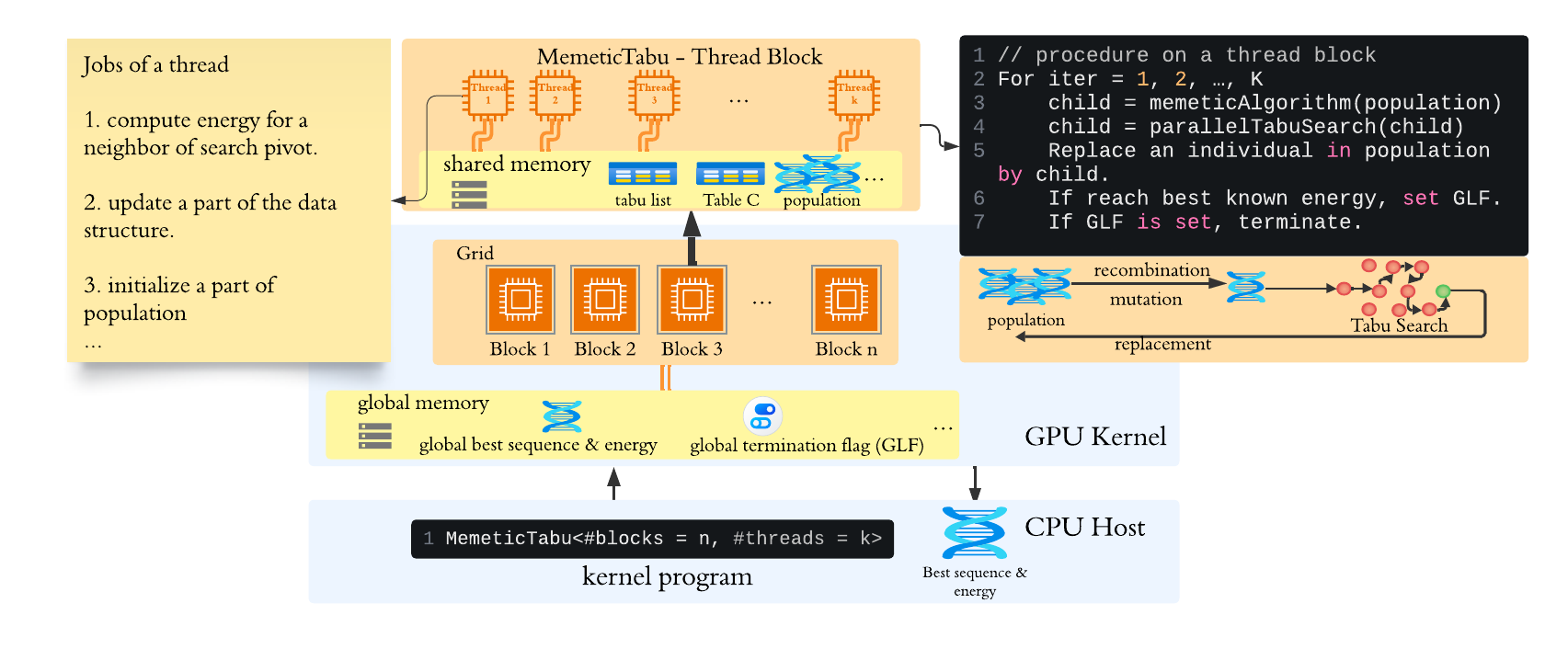}
    \caption{
    Architecture of our Memetic Algorithm with Tabu Search approach on GPUs. The CPU host launches a single kernel MemeticTabu once, avoiding data transportation and host-device switching. In the kernel, each thread block runs a replica of memetic tabu search algorithm.  The fast-but-small shared memory in each block, which can be accessed by all threads, is fully exploited by storing data structures. In each block, the computationally heavy steps are parallelized by multiple threads in a block.  Early termination is enabled by storing the global termination flag (GLF) as well as the best sequence and its energy, in the global memory of GPU. 
    }
    \label{fig:gpuarch}
\end{figure*}
Although the memetic-tabu algorithm described in Section \ref{sec:methodology} exhibits competitive exponential scaling \cite{gallardo2007memetic}, previous implementations have been inefficient for two reasons. First, the time needed to reach the target value of a single run depends heavily on the initialization. To reach the target value quickly, it is ideal to run a number of replicas with different random seeds simultaneously, which is challenging for CPU implementations due to the limitations in the desired number of cores. Second, previous implementations have not parallelized any part of the computationally expensive components, such as neighborhood energy checks, and update of the data structure.

In this work, we adapt the memetic-tabu algorithm to be massively parallelized, fully executed on GPUs. The parallelization is implemented at two distinct levels, which has recently been utilized for solving combinatorial optimization problems via heuristic search \cite{cen2025massively,kyrillidis2021continuous}.
\begin{enumerate}
    \item \textit{block-level} parallelism, where multiple algorithm replicas are executed concurrently, each initialized with different random seeds, across the thread blocks of a GPU;
    \item \textit{thread-level} parallelism, where computationally intensive steps of the algorithm within each replica are executed in parallel by multiple threads within a single thread block.
\end{enumerate}
We use compact data structures such as bit vectors to fit all data structures in limited but fast shared memory to exploit its efficiency compared against GPU global memory. A global termination flag is stored in the global memory for early termination when any block reaches the target value. By this architecture, thousands of thread blocks can be activated simultaneously, quickly exploring the search space with different initializations.  Our framework for parallelizing memetic-tabu algorithm on a GPU is illustrated in Figure \ref{fig:gpuarch}.

\subsection{All-in-GPU: Block-level and Thread-Level Parallelism}

\begin{figure}[ht]\begin{algorithm}[H]
\caption{\texttt{Memetic-Tabu-GPU}$\langle$\texttt{blocks}, \texttt{threadsPerBlock}$\rangle$(N, \texttt{targetE})}\label{algo:memetabuGPU}
\begin{algorithmic}[1]
\STATE \textbf{\#For each block do in parallel}
\STATE \texttt{population} $\gets$ $K$ random binary sequences of length $N$
\STATE \texttt{bestSeq} $\gets$ sequence with lowest energy from \texttt{population}
\WHILE{\texttt{E}(\texttt{bestSeq}) > \texttt{targetE}}
    \IF{$\texttt{randReal}(0,1) < p_{comb}$}
         \STATE $\texttt{parent}_1, \texttt{parent}_2 \gets \texttt{selectParents}(\texttt{population})$.
        \STATE $\texttt{child} \gets \texttt{Combine}(\texttt{parent}_1, \texttt{parent}_2)$
    \ELSE
        \STATE $\texttt{child} \gets$ a random individual from \texttt{population}
    \ENDIF
    \STATE $\texttt{Mutate}(\texttt{child}, p_{mutate})$
    \STATE $\texttt{child} \gets \texttt{Tabu-Search}(\texttt{child})$
    \STATE update \texttt{bestSeq} to \texttt{child} if \texttt{E}(\texttt{bestSeq}) $>$ \texttt{E}(\texttt{child})
    \STATE replace a random individual in \texttt{population} by \texttt{child}
    \STATE \textbf{if} \texttt{E}(\texttt{child}) $\leq$ \texttt{targetE} \textbf{then} \texttt{GLF} $\gets$ \texttt{True}
    \STATE \textbf{if} \texttt{GLF} \textbf{then return}
\ENDWHILE
\end{algorithmic}
\label{algo:memetabuGPU}
\end{algorithm}\end{figure}
Most of the running time in Algorithm \ref{algo:memetabuGPU} is spent, however, in tabu search, where energy evaluations are frequently needed.  We parallelize tabu search by multiple threads in a block as shown in Algorithm \ref{alg:tsgpu}. Specifically, the parallelization for neighborhood energy checking and data structure updating theoretically reduce the running time  by a factor of \texttt{threadsPerBlock}. 
Our design of a GPU implementation includes two levels of parallelism. At the block level,  multiple replicas are run with different random initializations in each thread block of a GPU to increase the probability of reaching the target value. At the thread level, several key steps in the algorithm are parallelized by multiple threads in each replica. All the computational work is done in GPU, which avoids the overhead of frequent host-device switch and data transportation.

Each block of the GPU runs a replica of memetic-tabu algorithm, shown in Algorithm \ref{algo:memetabuGPU}, independently with a different random seed. Therefore, a large number of blocks in GPUs can explore a  diverse  set of regions in the solution space simultaneously. A global termination flag (\texttt{GLF}) is placed in the global memory of GPU which can be accessed by all blocks. When a block reaches the target value, it sets \texttt{GLF} so that all blocks can shut down timely.

\begin{figure}\begin{algorithm}[H]
\caption{\texttt{Tabu-Search}(\texttt{seq})<threads>}\label{alg:tsgpu}
\begin{algorithmic}[1]
\STATE initialize \texttt{tabuList}$[N]=0$ and \texttt{bestSeqTS}=\texttt{seq}
\STATE \texttt{maxIter} $\gets \texttt{randomInt}(0,N) + \floor{\frac{N}{2}}$
\STATE \texttt{pivot} $\gets$ \texttt{seq}
\FOR{$t =1,2,\cdots, \texttt{maxIter}$}
\STATE \textbf{\#For each thread do in parallel\{}
    \STATE  $
i \gets \mathop{\arg\min}\limits_{i=1,\ldots,N, \texttt{tabuList}[i]<t} \texttt{E}(\texttt{flip}(\texttt{pivot}, i))
$
    \STATE /* Evaluate energy of all $N$ neighborhoods of \texttt{pivot}. Randomly choose one if there are multiple */ \textbf{\}}
    \STATE $\texttt{pivot}\gets\texttt{flip}(\texttt{pivot,i})$ 
    \STATE \texttt{tabuList}[i] = $t+ \texttt{randomInt(minTabu, maxTabu)}$
    \STATE update \texttt{bestSeqTS} to \texttt{pivot} if \texttt{E}(\texttt{bestSeqTS}) $>$ \texttt{E}(\texttt{\texttt{pivot}})
\ENDFOR
\RETURN \texttt{bestSeqTS}
\end{algorithmic}
\end{algorithm}\end{figure}

\subsection{Shared Memory Exploitation and Compact Data Structures}
The shared memory of a GPU, accessible by all threads of a block, can be $100+$ times faster than the global memory though with considerably limited size \cite{guide2020cuda}.  To exploit the efficiency of shared memory in our design, all data used in the algorithm including \texttt{population}, \texttt{tabulist}, \texttt{vectorC} and \texttt{tableC} are compactly stored in shared memory.  Nevertheless, to maximize the number of active blocks running simultaneously, each replica can only use a limited amount of shared memory. For instance, on a Nvidia-A100 GPU, each block should use no more than $\texttt{maxSharedMemPerSM/maxBlocksPerSM} =  164 \text{KB}/ 32 \approx$ 5 KB shared memory, which makes fitting all data structures in the shared memory a challenging task.

Given the limited amount of memory, we optimize the storage of all binary-valued data structures \texttt{population} and \texttt{tableC}, by employing compact bit-vector representations implemented through integer arrays. This approach results in an eightfold reduction in memory usage compared to previous implementations, thereby significantly enhancing the efficiency of data storage and management within the algorithm.
Moreover, the matrix \texttt{tableC} is upper-left triangular so only the non-zero entries need to be stored, saving another $\binom{N}{2}$ bits. In our implementation, the shared memory size 5 KB suffices for $N\leqslant 187$ for population size $K=100$,  while directly adapting the previous implementation makes the memory size of 5 KB insufficient even for $N\leqslant 11$.

\setlength{\tabcolsep}{6pt}
\begin{table*}[t!]
    \centering
    \begin{tabular}{>{\bfseries}c c c c c >{\centering\arraybackslash}p{6cm} c}
        \toprule
\textbf{N} & \textbf{Old E} & \textbf{New E} & \textbf{Old MF} & \textbf{New MF} & \textbf{New Sequences $S$} &
$d(S)$
\\
        \midrule
        \multicolumn{7}{c}{\textbf{Even Sequences}} \\
        \midrule
92 & 498 & 490 & 8.50 & 8.64 & EE01C0E77667DD34DAE94B5 & 12 \\ 
 &  &  &  &  & BB5495B2233288618FBC1E0 & 9 \\
98 & 545 & 529 & 8.81 & 9.08 & 993A76393CDB4BCF78FF00AA8 & 13\\ 
104 & 612 & 600 & 8.84 & 9.01 & 0F80F3C20AAB1844295B6ED666 & 12 \\ 
106 & 701 & 645 & 8.01 & 8.71 & 696D27CA748C66071EAFDD1177C & 14 \\ 
108 & 702 & 670 & 8.31 & 8.70 & F2401F3D9FF1CF46D58D96AA959 & 14 \\ 
110 & 723 & 707 & 8.37 & 8.56 & 33F83762F128DF55064B5B7BE79C & 16 \\ 
112 & 788 & 728 & 7.96 & 8.62 & A0496A493FAECCC8AFC3D50E738F & 13 \\ 
114 & 817 & 777 & 7.95 & 8.36 & B6C3648DB19C8C387A9EAA8578000 & 15 \\ 
116 & 814 & 802 & 8.27 & 8.39 & 52DF72096B92DCC87407044C8E1D5 & 14 \\ 
118 & 847 & 807 & 8.22 & 8.63 & E003FB9F87C674E5CD6D34CCB2AD54 & 15 \\ 
120 & 872 & 824 & 8.26 & 8.74 & 84AD68060864EE7989E1A6C0BA7551 & 18 \\ 
        \midrule
        \multicolumn{6}{c}{\textbf{Odd Sequences}} \\
        \midrule
99 & 577 & 553 & 8.49 & 8.86 & 0CF30C003783CBCC8DA92AAD4 & 15 \\ 
107 & 677 & 645 & 8.46 & 8.88 & E3C5CA9D21C2792DED3FC88AEEE & 17 \\ 
109 & 662 & 658 & 8.97 & 9.03 & DA62534CFBB114B94A3EBE3BE1F8 & 17 \\ 
113 & 752 & 736 & 8.49 & 8.67 & 5C3E8E2CC45A6DD1B794047295800 & 19 \\ 
119 & 835 & 787 & 8.48 & 9.00 & 81F0200DDD2B8C4E654E18ACA9A69A & 20 \\ 
        \bottomrule
    \end{tabular}

\caption{New Results for the LABS Problem with old and new energy (E), Merit Factors (MF), sequences in hexadecimal encoding (zeros are attached to the end of the sequence to make their lengths multipliers of $4$) and deviation from skew-symmetry ($d$). All sequences we found are with high $d(S)(\geqslant 9)$, indicating they are not likely to be found by skew-symmetry-based methods. For $N=92$ two non-equivalent sequences are discovered.}
\label{tab:labs_results}
\end{table*}

\section{Results} \label{sec:results}
\subsection{Experiment Setup}
We implement our approach in C++ and CUDA. The experiments are conducted in a single Nvidia-A100 GPU with 6912 CUDA cores, 80 GB memory and 164 KB shared memory per streaming processor (SM). We dynamically assign the number of blocks and number of threads per block according to the problem size $N$. As it is a common practice to set \texttt{threadsPerBlock} as a multiplier of number of threads in a GPU wrap (32), we set \texttt{threadsPerBlock} to the smallest multiplier of $32$ that is larger than $N$. The number of blocks is set to 
\begin{equation*}
\small
    \texttt{blocks}=\texttt{\#SMs} \times \min(\texttt{maxBlocksperSM}, \frac{\texttt{maxTreadsPerSM}}{\texttt{threadsPerBlock}}),
\end{equation*}
to maximize the number of active blocks under the restrictions of the specific GPU configurations.

The stopping criterion is set to be reaching the previous best-known results. We also run the implementation of Ref.~\cite{gallardo2007memetic} on an AWS c5.24xlarge CPU node with 48 physical cores at 3.056 GHZ for a comparison with multicore CPU implementations.

\subsection{Showcasing New MF Values}
The optimal solution for the LABS problem with $N > 66$ has remained unknown, primarily due to the non-scalability of the global branch-and-bound method \cite{Packebusch2016}. Notably, the best-known solutions for problem sizes up to $N \leqslant 120$, derived through meta-heuristic approaches, have not seen improvements since 2020 \cite{OPUS2-git_labs-Boskovic}. However, our implementation of a GPU-parallelized memetic-tabu search has demonstrated advancements in this space for $92 \leqslant N \leqslant 120$ by obtaining sequences with better merit factors compared to the previously best-known results, as detailed in Table \ref{tab:labs_results} and Figure \ref{fig:improvement}.
 These improved results were achieved by executing our implementation on a single Nvidia-A100 GPU with 96 hours, for odd  $ N \leqslant 109$ and even $N \leqslant 118$.

For even sequences within the range of $N = 92$ to $N = 120$, our study has yielded improved results for eleven problem sizes: specifically, $N = 92, 98, 104, 106, 108, 110, 112, 114, 116, 118, 120$. For problem sizes $N = 94, 96, 100, 102$, the sequences generated by our implementation exhibit MF values that are consistent with the previously reported best-known results~\cite{OPUS2-git_labs-Boskovic}. Additionally, our implementation has identified five novel results five odd sequences at $N = 99, 107, 109, 113, 119$. Notably, while the prior best-known results for these sequences were associated with optimal skew-symmetric sequences \cite{Packebusch2016}, our findings indicate that the optimal objective values for these problems do not coincide with the optimal skew-symmetric values. 

\subsection{Scaling and Runtime}

\begin{figure}[h!]
    \centering
    \includegraphics[scale=0.3]{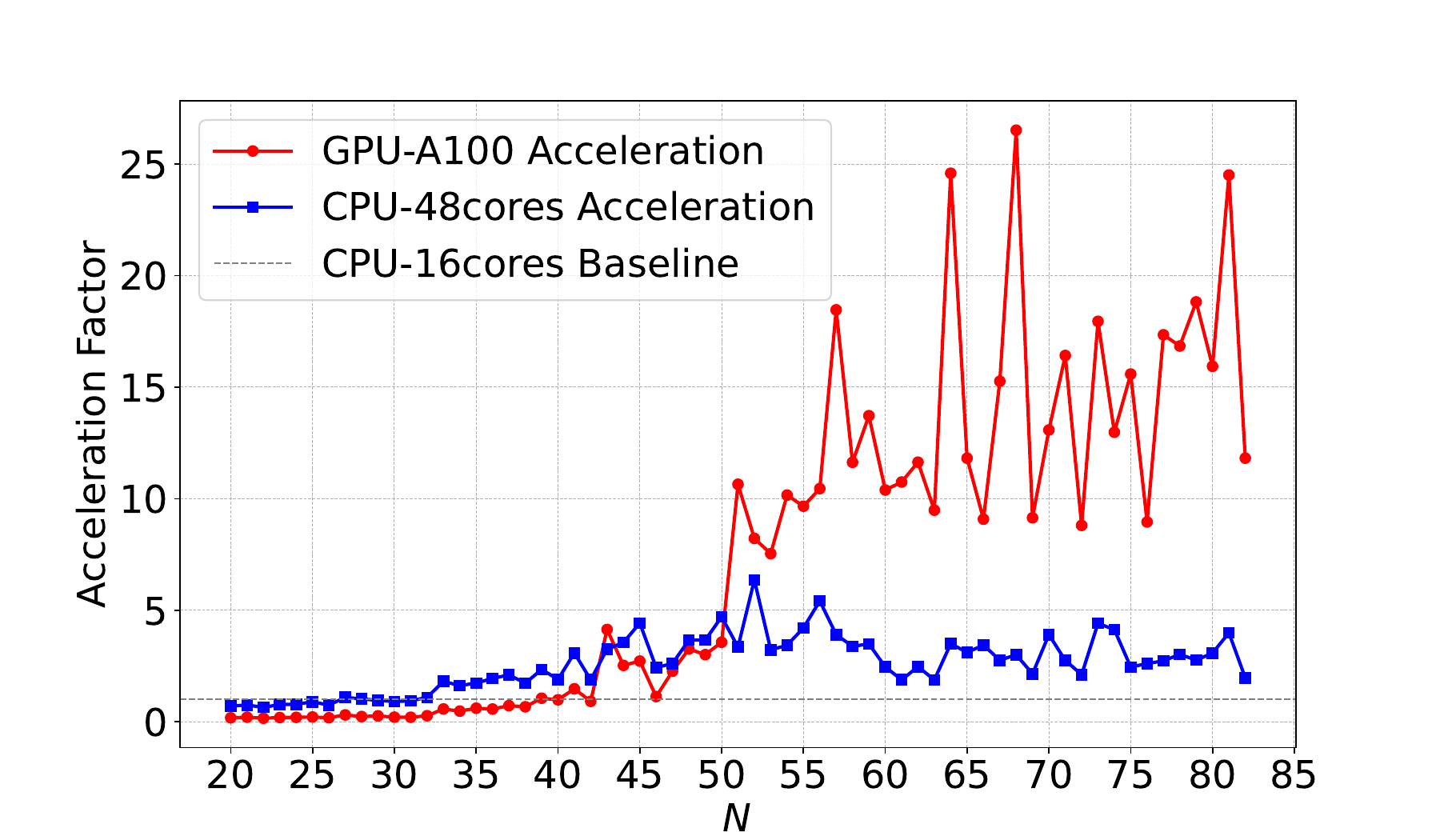}
    \caption{
    Acceleration of our GPU implementation against previous version on multiple CPU cores for time-to-solution (TTS). The most significant speed-up is 26.5$\times$ on $N=68$, where the GPU implementation takes 5.15 seconds and the CPU version needs 136.64 and 45.43 seconds for 16 and 48 cores respectively. 
    }
    \label{fig:acceleration}
\end{figure}

\begin{figure}[htbp]
    \centering
    \includegraphics[width=0.95\linewidth]{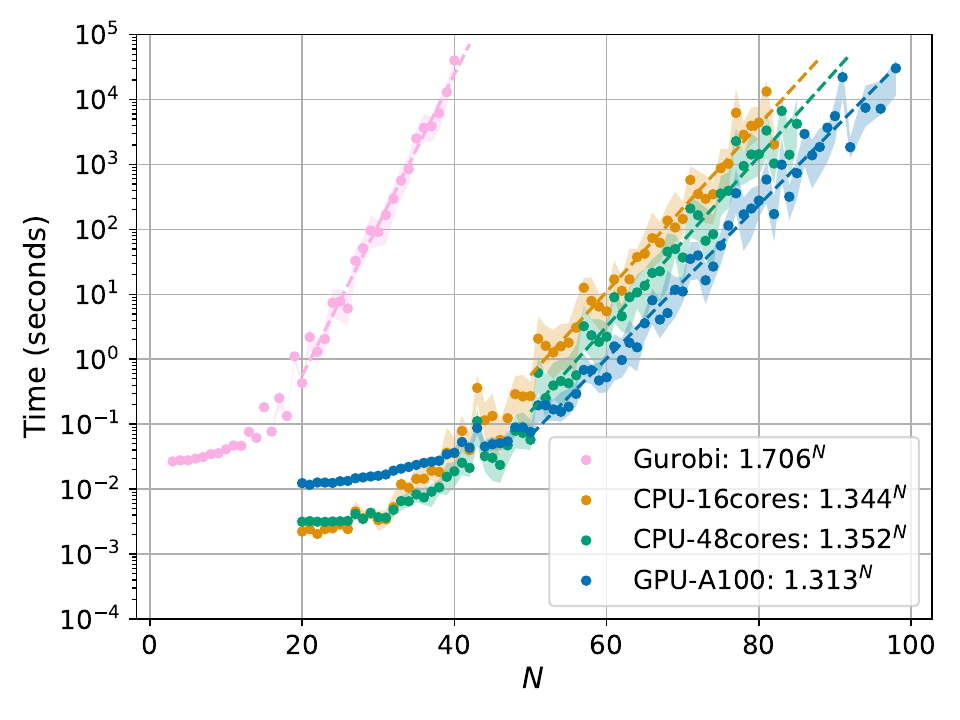}
    \caption{
    Median time to target energies with the error band representing the intervals from Q1 (25\% percentile) to Q3 (75\% percentile).
    The scalings are from exponential fits with (1) $N_{\mathrm{min}}^{\mathrm{fit}} = 20$ for Gurobi; (2) $N_{\mathrm{min}}^{\mathrm{fit}} = 50$ for memetic-tabu GPU-A100, CPU-16cores, and CPU-48cores. 
    }
    \label{fig:runtime_various_machines}
\end{figure}

Our implementation of the memetic algorithm with tabu search on a GPU was tested on the LABS problem for all problem sizes $N\leqslant 120$.  As illustrated in Figure~\ref{fig:runtime_various_machines}, the GPU-based memetic tabu search maintains the same scaling factor as its analogous CPU implementation. Notably, Figure \ref{fig:acceleration} demonstrates the constant factor speedup achieved by the GPU implementation. Specifically, our approach delivers an 8- to 26-fold acceleration for problem sizes ranging from 55 to 83 when utilizing an Nvidia-A100 GPU, compared to the analogous implementation on a 16-core CPU.

\begin{figure}[htbp]
    \centering
    \includegraphics[width=0.95\linewidth]{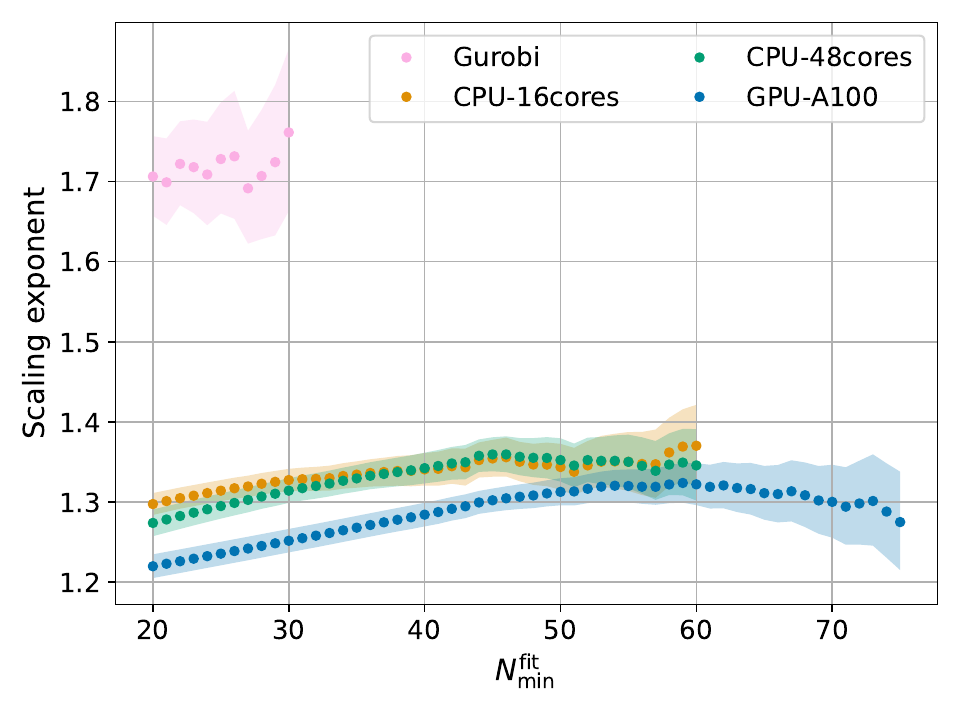}
    \caption{
    The scaling factor $b$ from the $a \times b^N$ fit with the error band representing the 95\% CI.
    }
    \label{fig:scalings_various_machines}
\end{figure}

\setlength{\tabcolsep}{6pt}
\begin{table}[!htbp]
    \centering
    \begin{tabular}{ccc} \toprule
    $N_{\mathrm{min}}^{\mathrm{fit}}$ & $a$ (95\% CI) & $b$ (95\% CI) \\
    \midrule
    20 & 2.05e-05 (9.71e-06, 4.34e-05) & 1.220 (1.205, 1.235) \\
    25 & 8.22e-06 (3.86e-06, 1.75e-05) & 1.235 (1.221, 1.250) \\
    30 & 3.20e-06 (1.49e-06, 6.88e-06) & 1.252 (1.237, 1.266) \\
    35 & 1.22e-06 (5.59e-07, 2.65e-06) & 1.268 (1.253, 1.283) \\
    40 & 4.60e-07 (2.06e-07, 1.03e-06) & 1.284 (1.269, 1.299) \\
    45 & 1.56e-07 (7.01e-08, 3.49e-07) & 1.302 (1.288, 1.317) \\
    50 & 8.23e-08 (3.18e-08, 2.13e-07) & 1.313 (1.296, 1.330) \\
    55 & 5.24e-08 (1.61e-08, 1.71e-07) & 1.320 (1.300, 1.341) \\
    60 & 4.61e-08 (9.76e-09, 2.18e-07) & 1.322 (1.296, 1.348) \\
    65 & 9.37e-08 (1.18e-08, 7.42e-07) & 1.311 (1.278, 1.345) \\
    70 & 1.93e-07 (1.05e-08, 3.55e-06) & 1.300 (1.255, 1.346) \\
    \bottomrule
    \end{tabular}
    \caption{Fit results with varying $N_{\mathrm{min}}^{\mathrm{fit}}$ of the median runtime with fit function $a \times b^N$ (in seconds). Numbers in parentheses are 95\% confidence intervals.}
    \label{tab:fit}
\end{table}

Figure~\ref{fig:runtime_various_machines} shows the time-to-solution (TTS) scaling of our GPU implementation of the memetic-tabu algorithm and comparing it with the analogous 16-core and 48-core CPU implementation. Further, as a reference, we include the TTS scaling of Gurobi, the global optimization solver based on branch-and-bound~\cite{Gurobi}. The experiments for Gurobi were conducted with Gurobi version 11.0.3 on 16 threads on a AMD CPU with 48 physical cores and 96 logical cores. In Gurobi, the LABS problem is encoded as a quadratic constrained binary optimization problem with auxiliary variables~\cite{kratica2012mixed}. Although Gurobi is a global optimization solver, in our experiments we set the stopping criterion to reaching the previous best-known energy, instead of proving the global optimality.

For the TTS scaling, we fit the median runtime by the function $a\times b^N$ for $N$ in intervals $[N_{\mathrm{min}}^{\mathrm{fit}}, N_{\mathrm{max}}^{\mathrm{fit}}]$ where $N_{\mathrm{min}}^{\mathrm{fit}}$ is the starting point in the fit as a variable. $N_{\mathrm{max}}^{\mathrm{fit}}$ is the end point in the fit, taken as $40$ for Gurobi, $82$ for memetic-tabu CPU-16cores, $85$ for CPU-64cores and $98$ for GPU-A100. A linear fit is conducted with ordinary least squares for the logarithm of median runtime versus $N$, and then the logarithm gets converted back to times in seconds. The fit function values $a$ and $b$, along with the range within $95\%$ confidence interval (CI), for our GPU parallelized memetic tabu implementation is reported in Table~\ref{tab:fit}. 

Figure~\ref{fig:scalings_various_machines} compares the scaling factor, $b$, for the three implementations of memetic tabu and the Gurobi solver. We note that memetic tabu implementation runtime for $N \lesssim 40$ is dominated by the kernel launching overhead and problem instance construction, so the exponential fit with these points may lead to artificially lower scaling factors. Consequently, we report the TTS of memetic-tabu with $N_{\mathrm{min}}^{\mathrm{fit}} = 50$ in Figure~\ref{fig:runtime_various_machines}. Our GPU implementation scaling factor, as reported in Table~\ref{tab:fit}, aligns with Ref.~\cite{gallardo2007memetic} with their reported scalings $1.34^N$, and $1.35^N$ reported in Refs.~\cite{Bokovi2017,Shaydulin2024}. Further, the 95\%-CI of the Gurobi scaling has overlap with that of the time-to-solution (TTS) scaling of Gurobi reported in Ref.~\cite{Shaydulin2024}.

\subsection{Deviation from Skew-Symmetry}

{In Appendix~\ref{sec:analysis_deviation_skew_symmetry}, we introduce $d(S)$, the deviation from skew-symmetry for sequence $S$.
The definition of $d(S)$ applies for both odd and even sequences. We calculate $d(S)$ for all our newly found sequences and show the results in Table~\ref{tab:labs_results}. The $d(S)$ of our sequences is at least $9$, which are considerably greater than a typical sequence found by the skew-symmetry-based method of Ref.~\cite{pvsenivcnik2024dual}, giving evidence that our sequences are not likely to be found by skew-symmetry-based methods.
In Ref.~\cite{Prestwich2013} all optimal skew-symmetric sequences for odd $N\leqslant 119$ are found and proved optimal via branch-and-bound. The previous best results for $N=99,107,109, 113, 119$ are given by optimal skew-symmetric sequences before this work. Our non-skew-symmetric results for $N=99, 107,109, 113, 119$ further confirm the global sub-optimality of optimal skew-symmetric sequences.}

\section{Discussion} \label{sec:discussion}

In this work, we present a framework for the massive parallelization of the memetic algorithm with tabu search, specifically tailored for the LABS problem on a single Nvidia-A100 GPU. Our implementation leverages a carefully crafted design of block-level and thread-level parallelism, further optimized by utilizing the fast shared memory available among blocks. By encoding the algorithm's data storage in a binary-valued format, we achieve optimal GPU utilization, thereby enhancing computational efficiency.

Using LABS as a testbed, we showcase the efficacy of our implementation in rapidly finding the best-known solution up to problem size $120$. This is illustrated by up to a 26-fold speedup of our GPU implementation over the analogous memetic tabu implementation on a 16-core GPU. Additionally, Our implementation exhibits a considerable scaling advantage over the global branch-and-bound solver of Gurobi for problem sizes $N\leqslant 91$. Further, our optimized implementation has also allowed us to improve upon the best-known LABS objective value for sixteen different problem sizes from $92\leqslant N \leqslant 120$. 

Among the newly identified sequences that exhibit improvements over previously established LABS objective values, five belong to odd problem sizes, specifically $N = 99,107,109,113, 119$. This is of particular importance, since odd-sized LABS problems exhibit the skew-symmetry property which quadratically reduces the search space. Historically, leveraging this property, prior methods have reported optimal skew-symmetric LABS objective values for these problem sizes, which were also the best known LABS objective values. Our improved results for these problem sizes highlight the potential limitations of solely utilizing the skew-symmetry property to constrain the search space since this approach may inadvertently increase susceptibility to local optima. This suggests that a general-purpose method performing unrestricted search across the entire space remains advantageous, particularly for complex search landscape of LABS.

\section*{Acknowledgments}
We thank our colleagues at the Global Technology Applied Research center of JPMorganChase for support and helpful feedback. Special thanks to Ruslan Shaydulin, Yue Sun, Atithi Acharya, Rudy Raymond, and Tyler Chen for their valuable discussions regarding the manuscript, to Jacob Albus for technical support in setting up Gurobi experiments, and to Pragna Subrahmanya and Sriram Gunja Yechan for technical support around GPU implementation. 

\appendix

\section{Analysis of deviation from skew-symmetry for solutions}\label{sec:analysis_deviation_skew_symmetry}

For odd $N = 2 n - 1$, skew-symmetry is defined as the condition $s_{n+i} = (-1)^i s_{n-i}$ for all possible pairs, \textit{i.e.}, $i = 1, 2, \cdots, n - 1$. The number of pairs do not satisfy the skew-symmetric condition can be used as a metric for how distant a sequence is from skew-symmetry. Formally, the number of non-skew-symmetric pairs for a sequence $S = s_1 s_2 \cdots s_N$ can be calculated as
\begin{equation}
    n_\mathrm{non-skew} (S) =  \frac{1}{2} \sum_{i = 1}^{n-1} |{s_{n+i} - (-1)^i s_{n-i}}|
    .
\end{equation}

For even-length sequences, Ref.~\cite{dimitrov2022new} introduces sequence operators that append or delete elements from the beginning or the end of sequences, so that even-length solutions can be built based on odd-length skew-symmetric solutions, and the deviation from skew-symmetry of an even-length sequence can also be analyzed. Moreover, Ref.~\cite{pvsenivcnik2024dual} uses the sequence operators and also combines with the rotation defines as
\begin{equation}
    \mathrm{Rot} (s_1 s_2 \cdots s_N; i) = s_{i+1} s_{i+2} \cdots s_N s_1 \cdots s_i
\end{equation}
which moves the first $i$  elements to the end of the sequence.

``Compositions of sequence operators (appending and deleting at the beginning and/or the end) and rotation'' are equivalent to the ``compositions of insertion and deletion at arbitrary locations and rotation''. Therefore, we define the insertion and deletion operators at arbitrary locations as
\begin{equation}
    \mathrm{Ins} (s_1 s_2 \cdots s_N; i, \xi) = s_1 s_2 \cdots s_i \xi s_{i+1} \cdots s_N,
\end{equation}
\begin{equation}
    \mathrm{Del} (s_1 s_2 \cdots s_N; i) = s_1 s_2 \cdots s_{i-1} s_{i+1} \cdots s_N,
\end{equation}
where $\xi \in \{-1, 1\}$. Allowing one time of insertion and deletion, we define the metric for the deviation of skew-symmetry as
\begin{widetext}
\begin{equation}
    d (S) = \min \left\{ \min_{i, r, \xi} n_{\mathrm{non-skew}} ( \mathrm{Ins} (\mathrm{Rot} (S; r); i, \xi) ), \min_{i, r} n_{\mathrm{non-skew}} ( \mathrm{Del} (\mathrm{Rot} (S; r); i) ) \right\}
    ,
\label{eq:n_non-skew_1-ins_del}
\end{equation}
\end{widetext}
as the minimum number of non-symmetric pairs with rotation and any insertion and deletion by one time. This gives some demonstration that the sequence may be difficult to obtain under the operations used by Ref.~\cite{pvsenivcnik2024dual}. In principle, multiple times of insertion and deletion can be allowed, but this would make the calculation of Eq.~\ref{eq:n_non-skew_1-ins_del} very expensive and eventually become equivalent to a complete search when the number of insertion/deletion becomes $\mathcal{O}(N)$.

We list the values of $d(S)$ for our newly found sequences in Table~\ref{tab:labs_results}. As a comparison, valid sequences from Ref.~\cite{pvsenivcnik2024dual} have $d(S) \leqslant 3$ for all odd $N$ except for $N=463$ and $d(S) \leqslant 3$ for all even $N$. This indicates it may be difficult for skew-symmetry-based methods to find better solutions, although we note that Ref.~\cite{pvsenivcnik2024dual} deals with $N \geqslant 450$ much greater than the regime we work with in this paper.

\bibliographystyle{plain}
\bibliography{refs}

\section*{Disclaimer}
This paper was prepared for informational purposes by the Global Technology Applied Research center of JPMorgan Chase \& Co. This paper is not a product of the Research Department of JPMorgan Chase \& Co. or its affiliates. Neither JPMorgan Chase \& Co. nor any of its affiliates makes any explicit or implied representation or warranty and none of them accept any liability in connection with this paper, including, without limitation, with respect to the completeness, accuracy, or reliability of the information contained herein and the potential legal, compliance, tax, or accounting effects thereof. This document is not intended as investment research or investment advice, or as a recommendation, offer, or solicitation for the purchase or sale of any security, financial instrument, financial product or service, or to be used in any way for evaluating the merits of participating in any transaction.
\end{document}